\documentclass[aps,prb,groupedaddress,twocolumn]{revtex4}
\usepackage{keyval} \usepackage{dcolumn}

\usepackage{times,xspace}
\usepackage{amsbsy,amssymb,bm,mathptmx}
\usepackage{graphicx,color,epsfig,rotate}
\usepackage{fancyhdr}
\usepackage{amsmath,amssymb,amsfonts,amsthm}
\usepackage{tabularx,siunitx}
\usepackage{booktabs}
\usepackage[]{units}

\usepackage[normalem]{ulem}
\setkeys{Gin}{width=0.90\columnwidth}
\graphicspath{{figs/}{:Figs:}} 

\renewcommand{\vec}[1]{\bm{#1}}
\newcommand{\mat}[1]{\overline{\bm{#1}}}

\newcommand{\comment}[1]{\textcolor{red}{#1}}
\renewcommand{\comment}[1]{\relax}
\newcommand{\tobedeleted}[1]{\textcolor{green}{\sout{#1}}}
\renewcommand{\tobedeleted}[1]{\relax}

\begin{document}

\title{The magnetic field generated by a charge in a uniaxial magnetoelectric material}

\author{M. Fechner }
\affiliation{Materials Theory, ETH Zurich, Wolfgang-Pauli-Strasse 27, CH-8093 Z\"{u}rich, Switzerland}
\author{I. E. Dzyaloshinskii}
\affiliation{School of Physical Sciences, University of California Irvine, Irvine, CA 92697, USA}
\author{N. A. Spaldin}
\affiliation{Materials Theory, ETH Zurich, Wolfgang-Pauli-Strasse 27, CH-8093 Z\"{u}rich, Switzerland}

\date{\today}

\begin{abstract}
We revisit the description of the magnetic field around antiferromagnetic magnetoelectrics in the context of recent developments regarding magnetoelectric monopoles. Using Maxwell's equations, we calculate the magnetic and electric fields associated with a free charge in a bulk uniaxial magnetoelectric, as well as in a finite sphere of magnetoelectric material. We show that a charge in the prototypical magnetoelectric Cr$_2$O$_3$, which is uniaxial with a diagonal magnetoelectric response, induces an internal magnetic field with both monopolar and quadrupolar components, but that only the quadrupolar contribution extends beyond the sample surface. We discuss the behavior of the external quadrupolar field and compare its magnitude to those of magnetic fields from other sources. 
\end{abstract}

\maketitle

A linear magnetoelectric is a material in which an applied magnetic field induces an electric polarization and an applied electric field induces a magnetization, with the size of the response proportional to the strength of the field:\cite{Fiebig:2005wa}
\begin{eqnarray}
P_i & = & \alpha_{ij} H_j \nonumber \\
M_i & = & \alpha_{ji} E_j \quad .
\end{eqnarray}
There has been increasing research interest in magnetoelectrics over the last fifteen years, motivated in part by the technological appeal of electric-field-controlled magnetism, as well as the intriguing mechanisms that allow a ferroic property to be modified other than by its conjugate field. In addition, it has been pointed out in the last months that certain symmetry classes of magnetoelectrics exhibit behaviors that can be described in terms of so-called magnetoelectric monopoles\cite{Spaldin:2013bo} leading to potentially new physics such as hidden orders and novel transport properties\cite{Khomskii:2013tx}.
 
The magnetoelectric tensor, $\alpha_{ij}$, is an axial second rank tensor, which is antisymmetric under both space and time inversion. As for any square tensor it can be decomposed into symmetric and antisymmetric parts:
\begin{equation}\label{eqn:def_alpha_general}
\mat{\alpha}=\mat{\alpha}_S+\mat{\alpha}_{AS} 
\end{equation}
with the anti-symmetric components indicating responses perpendicular to the external field. The form of $\mat{\alpha}$, of course reflects the symmetry of the system, and there exists a large number of uniaxial magnetoelectrics, including the prototype Cr$_2$O$_3$ \cite{Dzyaloshinskii:1960}, in which the response is purely diagonal in the basis of the crystallographic axes. For such materials $\mat{\alpha}_{AS}$ vanishes, and it is useful to reduce the symmetric part further into its isotropic diagonal and trace-free symmetric contributions: 
\begin{eqnarray}\label{eqn:def_alpha_diagonal}
\mat{\alpha}= \mat{\alpha}_S = \frac{1}{3}(2\alpha_\perp+\alpha_\| )\mathbb{I}+ \frac{\alpha_\perp-\alpha_\|}{3}\begin{bmatrix} 1&0&0\\ 0&1&0 \\ 0&0&-2  
\end{bmatrix} \quad .
\end{eqnarray}
Here $\alpha_\|$ and $\alpha_\perp$ are the responses parallel and perpendicular to the high symmetry axis respectively, and $\mathbb{I}$ is the unit matrix. When a correspondence is made between the second-order terms in the magnetic multipole expansion and the linear magnetoelectric response, it emerges that the first isotropic part corresponds to a ``magnetoelectric monopolar'' component, and the second traceless part to a ``magnetoelectric quadrupolar'' component \cite{Spaldin:2008dr,Spaldin:2013bo}.  (The antisymmetric part which is is zero in this case results from a magnetoelectric toroidal moment\cite{Ederer:2007je}.) A free electric charge in such a magnetoelectric should in turn generate a magnetic field that reflects these monopolar and quadrupolar contributions.

In this work we use Maxwell's equations to calculate the magnetic and electric fields associated with a free charge in a magnetoelectric material. We begin with the mathematically straightforward case of an isotropic magnetoelectric material and confirm that, as required by symmetry, it has a purely monopolar response. While isotropic magnetoelectrics have been discussed in the literature\cite{Hehl:2009cl}, none has been identified to date, as the requirement that time- and space-inversion symmetry be broken within a cubic symmetry is restrictive. Interest in them has been renewed recently in the context of their relationship to strong $Z_2$ topological insulators\cite{Coh:2013ec}, as well as their potentially novel transport properties\cite{Khomskii:2013tx}. We then proceed to the physically more abundant and mathematically more complex uniaxial case and evaluate the relative contributions of the monopolar and quadrupolar terms to the magnetic and electric fields. In both cases we obtain the solution for a free charge in an infinite magnetoelectric medium, as well as for the case of a finite spherical sample. Finally, based on our results we revisit literature magnetometry measurements of the magnetic field around Cr$_2$O$_3$\cite{Astrov:2013ua,Astrov:2013wu}. These were discussed previously in terms of the intrinsic quadrupolar response of an uncharged magnetoelectric\cite{Dzyaloshinskii:2002wf} as well as in terms of the surface magnetization caused by the antiferromagnetism \cite{Andreev:1996vi} and its coupling to the magnetoelectricity \cite{Belashchenko:2010ky}. We evaluate the relative magnitudes of all three contributions to the external magnetic field and discuss the implications.

\section{An isotropic magnetoelectric}
\begin{figure}
\centering
\def\svgwidth{\columnwidth}
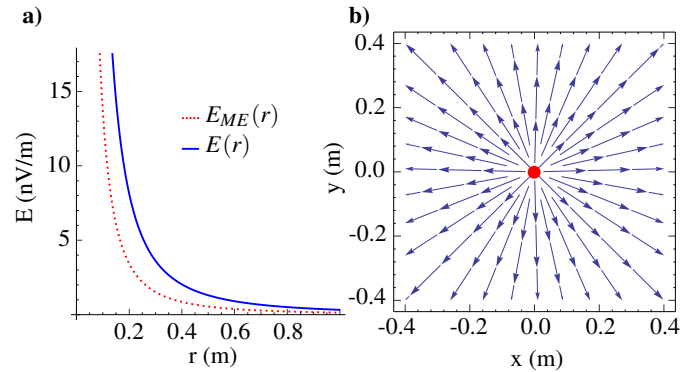
\caption{\label{fig1} (a) Electric field around a point charge in an isotropic dielectric (blue solid line) and in a magnetoelectric with $\alpha$ increased by a factor of 20,000 (red dotted line). The parameters used are the measured values for Cr$_2$O$_3$, averaged to give isotropic values. (b) Calculated magnetic field, $\vec{H}(\vec{r})$, generated by a point charge in an isotropic magnetoelectric in a cut through the a plane containing the charge. The arrows indicate the field orientation.}
\end{figure}

We provide a source of electric field within an isotropic magnetoelectric by introducing a point charge $\rho=q\delta(r)$. While in practice such a charge is likely to be an electron, here we treat the case of a spinless charge, and do not include in our analysis the reciprocal magnetoelectric response arising from an electronic spin magnetic moment. Since the material is isotropic, the $\mat{\alpha}$ tensor is given by
\begin{eqnarray*}
\mat{\alpha}= \begin{bmatrix} \alpha&0&0\\ 0&\alpha&0 \\ 0&0&\alpha \end{bmatrix}. 
\end{eqnarray*}
The magnetoelectric medium augments the displacement field and magnetic induction from their usual $\mat{\epsilon}\vec{E}$ and $\mat{\mu}\vec{H}$ to include the magnetoelectric cross terms: 
\begin{eqnarray}\label{eqn:def_DB}
\vec{D} & = & \mat{\epsilon}\vec{E}+\mat{\alpha} \vec{H}\nonumber\\
\vec{B} & = & \mat{\mu}\vec{H}+\mat{\alpha} \vec{E}.
\end{eqnarray}
Here $\mat{\epsilon}$ and $\mat{\mu}$ are the usual dielectric permittivity and magnetic permeability tensors, which have the same isotropic symmetry as the magnetoelectric tensor. Insertion of these field expressions into Maxwell's equations yields a system of coupled differential equations which we solve by introduction of a magnetic and electric potential\cite{Jackson:2007ub} (see appendix \ref{app:iso_inf}) to yield the fields:
\begin{eqnarray}\label{eqn:solution_isotropic_case}
\vec{E}(\vec{r})&=&\frac{ \mu}{4\pi(\epsilon\mu-\alpha^2)} \frac{q }{\vec{r^2}}\vec{e_r}\nonumber \\
 \vec{H}(\vec{r})&=-&\frac{ \alpha}{4\pi(\epsilon\mu-\alpha^2)} \frac{q}{\vec{r^2}}\vec{e_r} \quad,
\end{eqnarray}
where $\vec{e_r}$ is the unit vector in the radial direction. First we discuss the electric field. We see that its form is identical to that induced by a free charge in an ordinary dielectric  --  $\frac{1}{4 \pi \epsilon} \frac{q}{\vec{r^2}}$ -- except that the magnitude of the induced field is modified with an effective (lowered) dielectric constant $\epsilon_{eff}=(\epsilon-\alpha^2/\mu)$. To estimate the size of the lowering we use literature values for Cr$_2$O$_3$ (Table \ref{tab:parameters}) averaged as in Eqn. (\ref{eqn:def_alpha_diagonal}) to extract the isotropic part, and find that the change in relative permittivity is small, with $\Delta\epsilon_r=\alpha^2/(\mu \epsilon_0)\sim 1.5\times 10^{-8}$. In Fig.~\ref{fig1} (a) we compare the electric field of a pure dielectric with that of an isotropic magnetoelectric, with $\alpha$ 20,000 times larger than that of Cr$_2$O$_3$, for the true $\alpha$ the curves would be indistinguishable from each other. 

Eqns.~\ref{eqn:solution_isotropic_case} indicate that if $\alpha^2 \rightarrow \epsilon\mu$, both the 
electric and magnetic fields inside the medium would diverge causing a ``magnetoelectric catastrophe''. 
This is consistent with the inequality
\begin{equation*}
\alpha_{ij} \le \sqrt{\mu_{ii} \epsilon_{jj}}
\end{equation*}
required for thermodynamic stability \cite{ODell:1963ip} and the stricter constraint, derived using thermodynamic perturbation theory and requiring positive definiteness of the free energy with respect to external fields \cite{Brown:1968vl}, that $\alpha^2<\chi_m\chi_e$. Even close to a phase transition where $\alpha$ may diverge, either $\epsilon$ or $\mu$ will diverge simultaneously avoiding a magnetoelectric catastrophe. This scenario was discussed recently for a magnetoelectric-multiferroic phase transition using Landau mean-field theory \cite{Dzyaloshinskii:2011ib} and demonstrated numerically for the case of a strain-induced multiferroic phase transition\cite{Bousquet:2011eg} in CaMnO$_3$.

The magnetic field, $\vec{H}(\vec{r})$, shows a similar purely divergent radial dependence as the electric field. We illustrate this in Fig.~\ref{fig1} right panel, where the arrows represent the magnetic field lines. Note, however, that while $\vec{H}(\vec{r})$ diverges, $\vec{M}(\vec{r})=\alpha \vec{E}$ also diverges and exactly compensates the divergence in $\vec{H}$. As a result, $\nabla\cdot \vec{B}(\vec{r})=0$, Maxwell's equations are not violated and no ``true'' magnetic monopole is generated. 

\begin{table}
\begin{ruledtabular}
\begin{tabular}{c|rcc}
component & $\alpha$ (ps/m)  & $\epsilon_r$ ($\epsilon_0$)   & $\mu_r$ ($\mu_0$) \\ \hline
$\perp$   & 0.734 & 10.3   & 1.0014  \rule[-1ex]{0pt}{3.5ex}  \\
$\|$         & -0.233 & 10.9  & 1.0001  \rule[-1ex]{0pt}{3.5ex}  \\
\end{tabular}
\caption{Experimental values of $\alpha$, relative permittivity $\epsilon_r$ and relative permeability $\mu_r$ for Cr$_2$O$_3$\cite{Foner:1963vi,Wiegelmann:1994tq,LAL:1967fd}. $\epsilon_r$ is measured at room temperature, whereas $\mu_r$ and $\alpha$ are the low temperature (4 K) values. SI units are used throughout, so $\alpha$ has units of inverse velocity.}
\label{tab:parameters}
\end{ruledtabular}
\end{table}

\subsection{A finite sphere of an isotropic magnetoelectric}
\label{isotropicsphere}

\begin{figure}
\centering
\def\svgwidth{\columnwidth}
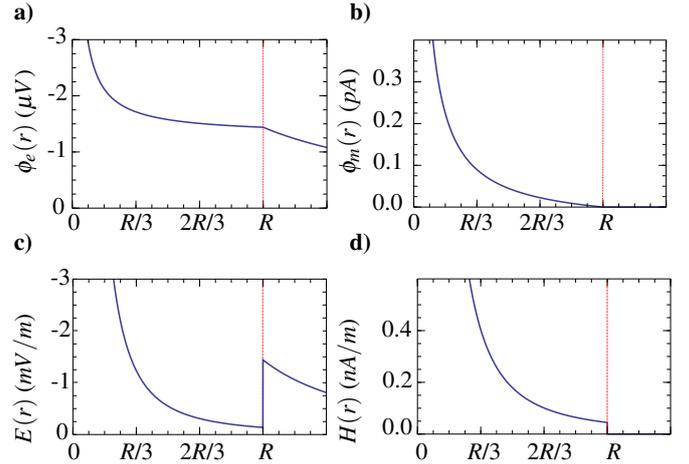
\caption{\label{fig2}(a) Electric and (b) magnetic potential as a function of distance from the center of the sphere. (c) Electric and (d) magnetic field. The sphere radius, $R$ = \unit[1]{mm}.}
\end{figure}

We now extend our discussion to the case of a sphere of an isotropic magnetoelectric with a charge at its center in a vacuum. The fields inside and outside of the sphere are subject to the boundary conditions: 
\begin{eqnarray}
\phi_{e,m}^i(\vec{R})&=&\phi_{e,m}^o(\vec{R})\nonumber\\
\vec{D}^i(\vec{R})\cdot \vec{n}&=&\vec{D}^o(\vec{R})\cdot \vec{n}\nonumber\\
\vec{B}^i(\vec{R})\cdot \vec{n}&=&\vec{B}^o(\vec{R})\cdot \vec{n}
\end{eqnarray}
where $\phi_{e,m}$ are the electric and magnetic potentials, the $o$ and $i$ superscripts indicate outside and inside of the sphere, respectively, $\vec{R}$ is the sphere radius, and $\vec{n}$ is a unit vector along the surface normal. $\vec{D}^i$ and $\vec{B}^i$ are given by the expressions of Eqns.~\ref{eqn:def_DB} including the magnetoelectric responses, whereas for the vacuum region the usual relations
\begin{eqnarray*}
\vec{D} & = & \epsilon_0\vec{E} \\
\vec{B} & = & \mu_0\vec{H}
\end{eqnarray*}
apply. We expand the potentials in a basis of Legendre polynomials, and since the system is radially symmetric all components apart from $l=0$ vanish. The full derivation is given in the appendix \ref{app:iso_boundary}, and our solutions are:
\begin{eqnarray}
\phi_e^i(\vec{r})&=&\frac{ \mu}{4\pi(\epsilon\mu-\alpha^2)} \frac{q}{\vec{r}}+\frac{(\alpha^2 +\mu(\epsilon-\epsilon_0))}{ (\alpha^2  -  \epsilon_0 \mu)}\frac{q}{4 \pi  \epsilon R}\nonumber\\
\phi_e^o(\vec{r})&=&\frac{1}{4\pi \epsilon_0}\frac{q}{r}\nonumber\\
\phi_m^i(\vec{r})&=&-\frac{\alpha}{4\pi(\epsilon\mu-\alpha^2)} \frac{q}{\vec{r}}+\frac{ \alpha}{ 4 \pi (\epsilon \mu-\alpha^2)}\frac{q}{ R}\nonumber\\
\phi_m^o(\vec{r})&=&0\;\;\; .
\end{eqnarray}
We plot our calculated potentials and fields, obtained using the parameters for Cr$_2$O$_3$ and a sphere radius, $\vec{R}$=\unit[1]mm, in Fig.~\ref{fig2}. 

Again we begin by discussing the electric field case. In Figs. \ref{fig2} (a) and (c) we show the radial components of the electric potential and the electric field inside and outside of the sphere. We see that the behavior is analogous to that of a charge at the center of a dielectric sphere, with the field falling off radially within the sphere, then more sharply in the vacuum region. There is a renormalization of the dielectric constant, however, from the magnetoelectric response. In Figs.~\ref{fig2} (b) and (d) we show the corresponding magnetic quantities. As in the case of the infinite medium, the $\vec{H}$ field diverges with radial distance from the charge, and $\vec{M}$ also diverges so that $\nabla\cdot\vec{B}$ is zero. Consequently the magnetic field vanishes at the boundary due to the zero magnetoelectric effect outside the sphere. As a result, all monopolar effects of the charge are constrained to within the sphere, and no magnetic field is detectable outside. 

Our main finding from this section, therefore, is that, while a charge in an isotropic magnetoelectric induces a monopole-like magnetization through the magnetoelectric effect, the magnetic field associated with this magnetization drops to zero at the boundary of a spherical sample. Thus, while the induced field might have a profound effect on the magnetotransport\cite{Khomskii:2013tx}, it cannot be directly detected outside of the sample. We point out, however, that any deviations from ideal sphericity in the sample will allow a non-zero, non-monopolar external field. 

\section{A uniaxial Magnetoelectric}
We now discuss the case of an anisotropic uniaxial magnetoelectric with a magnetoelectric tensor of the 
following form:
\begin{eqnarray*}
\mat{\alpha}= \begin{bmatrix} \alpha_\perp&0&0\\ 0&\alpha_\perp&0 \\ 0&0&\alpha_\| \end{bmatrix} .
\end{eqnarray*}
We obtain the fields in the same manner as in the isotropic case, working in this case in elliptical coordinates to more readily treat the different responses along the perpendicular and parallel axes. We note that, by symmetry, the $\mat{\mu}$ and $\mat{\epsilon}$ tensors have the same form as $\mat{\alpha}$, but the ratios of their parallel and perpendicular components are not necessarily equal. As a result fully analytical solutions are not accessible and the solutions -- given in appendix~\ref{app:uniaxial_inf} -- must be evaluated numerically. 

We begin by discussing the differences we expect from the isotropic case as a result of the anisotropy in each response tensor. In the isotropic case (Eqns. \ref{eqn:solution_isotropic_case}) we found that the electric field is described by an effective dielectric constant $\epsilon_{eff}=\epsilon-\alpha^2/\mu$. For Cr$_2$O$_3$, $\epsilon\sim$\unit[$10^{-12}$]{F/m} and $\alpha^2/\mu\sim$\unit[$10^{-19}$]{F/m}. Therefore we expect that the anisotropy of the electric field will be dominated by the anisotropy in the dielectric constant. For the magnetic field, the $\epsilon-\alpha^2/\mu$ term in the denominator is again dominated by $\epsilon$. However, there is also a prefactor of $\alpha/\mu$, and in the case of Cr$_2$O$_3$ the strongly anisotropic magnetoelectric tensor should determine the anisotropy in the magnetic field. Finally, we note that the magnetic field can become ``accidentally isotropic'' even when all tensors are anisotropic, if $\alpha_\perp/(\mu_\perp \epsilon_\perp)=\alpha_\|/(\mu_\| \epsilon\|)$. We will discuss an example of this behavior later. 

\begin{figure}
\centering
\def\svgwidth{1\columnwidth}
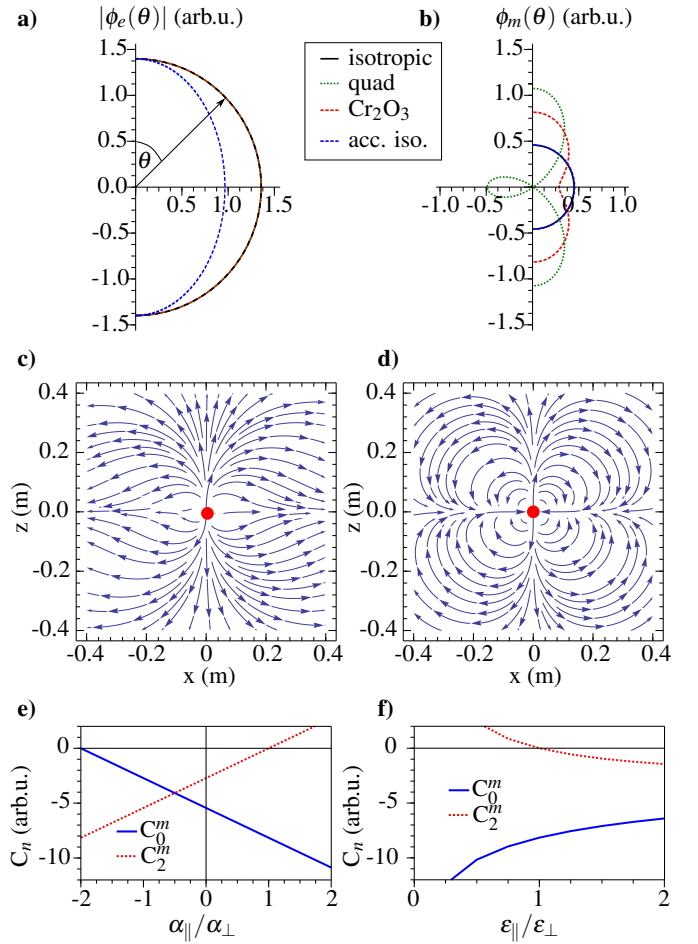
\caption{\label{fig3} (a) Electric and (b) magnetic potentials for four example sets of paremeters (for details see text.) (c) and (d): Magnetic fields $\vec{H}(\vec{r})$ in a cut through a plane containing the high symmetry axis, for Cr$_2$O$_3$ (c) and its quadrupolar-only contribution (d). The arrows indicate the field orientation. (e) and (f) Evolution of the expansion coefficents $C^m_0$ and $C^m_2$ corresponding to the monopolar and quadrupolar components of the magnetic potential, as a function of the ratio $\frac{\alpha_{\|}}{\alpha_{\perp}}$ and $\frac{\epsilon_{\|}}{\epsilon_{\perp}}$.}
\end{figure}

Now we look at our explicitly calculated azimuthal dependencies of the electric and magnetic potentials, which we show as polar plots in Figs.~\ref{fig3} (a) and (b) for four different sets of parameters: 
\begin{itemize}
\item The isotropic case obtained using averaged values for all three tensors (black lines), 
\item The actual measured parameters for Cr$_2$O$_3$ (red dashed lines), 
\item The measured Cr$_2$O$_3$ $\epsilon$ and $\mu$ values, but with only the tracefree quadrupolar part of the $\alpha$ tensor included (green dotted line), and 
\item The ``accidentally isotropic'' case, enforced by setting $(\epsilon,\alpha)_\perp=\frac{1}{2}(\epsilon,\alpha)_\|$ and $\mu_\|=\mu_\perp$ (blue dashed line).
\end{itemize}

We see in Fig.~\ref{fig3} (a) that the electric potentials for the true, isotropic, and quadrupolar-only cases lie exactly 
on top of each other, and are almost symmetrical, confirming our reasoning above that the asymmetry in $\epsilon$ (which is small in Cr$_2$O$_3$) determines that in the electric potential. The accidentally isotropic case has an elliptical-shaped electric potential, since we artifically set  $\epsilon_\perp=\frac{1}{2}\epsilon_\|$. 

In contrast there are large differences in the magnetic potential for the four different cases. As expected there is no angular dependence for the isotropic or accidentally isotropic cases, which overlie each other. The quadrupolar-only case (green line) changes sign with angle, consistent with its traceless magnetoelectric tensor. The true Cr$_2$O$_3$ potential (red line) is the sum of the isotropic and quadrupolar contributions and shows no sign change as the isotropic monopolar contribution dominates. In Figs.~\ref{fig3} (c) and (d) we show vector plots of the magnetic fields for the true Cr$_2$O$_3$ parameters and the purely quadrupolar case respectively. (d) shows clearly the field lines pointing outwards vertically, and inwards horizontally, in the manner of a quadrupole, whereas, because of the additional, larger, monopolar component in true Cr$_2$O$_3$, (c) shows outward pointing field lines in all directions. 

To quantify this division into monopolar and quadrupolar contributions, we expand the calculated electric and 
magnetic potentials in linear combinations of Legendre polynomials, where by symmetry only even terms are non-zero: 
\begin{equation}
\phi_{e,m}(\vec{r},\theta)=\frac{q}{4 \pi^2 r} \sum^{\infty}_{n=0} (4n+1) C^{e,m}_{2n} P_{2n}\left(cos(\theta)\right) \quad .
\end{equation}
Then each coefficient, $C_{2n}$, in the expansion indicates the contribution of the term with the corresponding azimuthal dependence, with $C_0$ representing the monopolar contribution, and $C_2$ the quadrupolar. Note that all contributions have the same $r^{-1}$ dependence, in contrast to the usual multipole expansion in which a multipole of order $n$ decays as $r^{-(n+1)}$, because of the infinite size of the system.

The relative contribution of each $C_{2n}$ is determined by the relative ratios of $\frac{\alpha_\|}{\alpha_\perp}$, $\frac{\epsilon_\|}{\epsilon_\perp}$ and $\frac{\mu_\|}{\mu_\perp}$. In Table \ref{tab:parameters} we already saw that the highest anisotropy occurs in the $\alpha$ tensor, with the dielectric constant and permeability being almost isotropic. (Note however that the magnetic {\it susceptibility} is strongly anisotropic with $\chi_\perp/\chi_\| \sim 15$ in Cr$_2$O$_3$\cite{Foner:1963vi}). In Fig.~\ref{fig3} (e) we plot the $C^m_0$ and $C^m_2$ expansion coefficients as a function of the ration $\frac{\alpha_\|}{\alpha_\perp}$, with the $\epsilon$ and $\mu$ tensors constrained to be purely isotropic. We find, as expected, that for $\frac{\alpha_\|}{\alpha_\perp}=-2$ the response is purely quadrupolar, whereas at $\frac{\alpha_\|}{\alpha_\perp}=1$ it is purely monopolar with a linear connection between the two limits. For comparison we show in (f) the same coefficients as a function of the corresponding asymmetry in $\epsilon$ and with an isotropic $\alpha$. For $\frac{\epsilon_\|}{\epsilon_\perp}=1$ a purely monopolar state found. 

Finally, we note that, in spite of the fact that materials with traceless magnetoelectric tensors exist -- an example is TbPO$_4$ which has one of the largest known magnetoelectric coefficients\cite{Rado/Ferrari:1972} -- charges in such materials do not generate purely quadrupolar electric and magnetic fields. This is because, while the magnetoelectric tensor can be traceless, the permittivity and permeability tensors can not, since the diagonal components of $\epsilon$ and $\mu$ are required for stability to be greater than 1. 

\subsection{A finite sphere of a uniaxial magnetoelectric}

Finally, we investigate the case of a sphere of uniaxial magnetoelectric surrounded by vacuum with a point charge at its center, and solve the corresponding equations subject to the same boundary conditions as in Section~\ref{isotropicsphere}. 
The solutions {\it inside} the sphere are identical to the previous solution for the infinite uniaxial case.  Each coefficient $C^{e,m}_{2n}$ then couples to a corresponding outside solution, with the exception of
$C^{m}_{0}$, which vanishes outside of the sphere as in the isotropic case, ensuring that $\nabla\cdot\vec{B}$ is always zero and Maxwell's equations are not violated. Higher order coefficients of $\vec{H}(\vec{r})$ and $\vec{E}(\vec{r})$ are, however, non-zero in the outside region, and in particular a finite quadrupolar magnetic field propagates beyond the sphere. The full derivation of the solution is presented in the appendix. 

\begin{figure}
\centering
\def\svgwidth{\columnwidth}
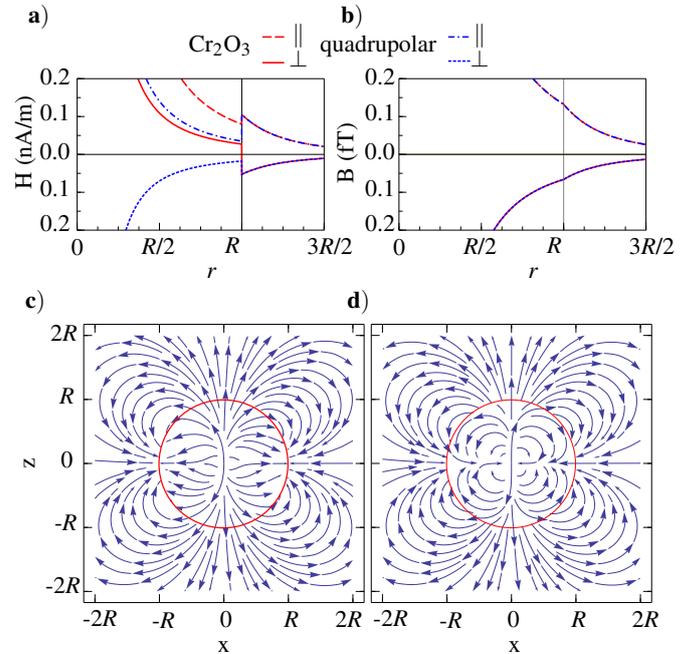
\caption{\label{fig4}(a) Magnetic field and (b) magnetic induction as a function of radial distance from the center of the sphere along the high symmetry $\perp$ and $\|$ directions. The blue lines corresponds to the quadrupolar-only response (dashed $\perp$, dot-dashed $\|$), and the red line shows the case of Cr$_2$O$_3$ (solid $\perp$, long-dashed $\|$). (c) and (d) show the magnetic fields in a slice through a plane containing the high-symmetry axis for Cr$_2$O$_3$ and the quadrupolar-only material respectively. The arrows show the orientation and magnitude of the field, with the magnitude weighted by the square of the distance from the charge.}
\end{figure}

As in the case for the infinite system, we find that the electric field for realistic parameters is close to isotropic therefore
we move directly to a discussion of the magnetic field. We consider two sets of parameters: The genuine Cr$_2$O$_3$ values, and
taking only the tracefree quadrupolar part of the $\mat{\alpha}$ tensor. In Fig.~\ref{fig4}, (a) we show the normal component of 
the magnetic field in both cases (dotted blue line: quadrupolar only, red line: full response)
along the two high symmetry directions ($\perp$ and $\|$). In contrast to the isotropic monopolar
case, the field does not vanish at the interface but extends into the vacuum area. Note that, since only the quadruoplar component
of the magnetic field extends out of the sphere, the fields are identical outside of the sphere in the two cases, in spite of the 
fact that they differ considerably within the sphere where the monopolar contribution can manifest. In Fig.~\ref{fig4} (b), we plot 
the normal component of the magnetic induction, $\vec{B}$, which is identical in the two cases. 

In Figs.~\ref{fig4} (c) and (d) we show the magnetic field lines in a cut through a plane containing the high symmetry axis for Cr$_2$O$_3$ 
and the purely quadrupolar case respectively. The red circle indicates the sphere boundary in both cases. Inside the sphere the solutions are identical to the infinite solutions shown in Fig.~\ref{fig3} (c) and (d). Again we
see that in the outside region again both fields are identical, as only the quadrupolar field extends out of the sphere. 
This indicates that magnetoelectrics with the same tracefree part of the the $\alpha$ tensor are indistinguishable from measurements
of their external magnetic field, provided that the $\epsilon$ and $\mu$ tensors are close to isotropic. 

\subsection{Comparison with other sources of magnetic field and with experiment}

Finally, we compare our calculated values for the magnetic field generated by a charge in a magnetoelectric
with that arising from two other sources: the intrinsic magnetoelectric response of an uncharged material,
and the surface moments from the antiferromagnetism. We then discuss the relevance of these possible contributions to the earlier measurements of Astrov and coworkers\cite{Astrov:2013ua}. 

First we point out that, in Fig. 4 (b) we see that an electron inside a magnetoelectric sphere of radius \unit[1]{mm} generates a magnetic induction of $\vec{B}_{ME}\sim$\unit[10$^{-16}$]{T} at the surface of the sphere via the magnetoelectric effect. We note that a single electron spin in a non-magnetoelectric sphere of the same dimension and permeability would generate a much smaller field of only \unit[10$^{-23}$]{T}, and so in this context, the magnetoelectric response should not be regarded as small. 

In Ref.~\onlinecite{Dzyaloshinskii:2002wf} we showed that the intrinsic structure and magnetic order in Cr$_2$O$_3$ give rise to an external quadrupole field even without an internal charge. The order of magnitude of this field at the surface is $\frac{\mu_B}{a V^{1/3}}$, with $a$ an atomic distance and $V$ the volume of the sample. For the spherical sample with radius \unit[1]{mm} that we have considered here, explicit calculation of the surface magnetic induction from this contribution gives $\vec{B}_{int}=$\unit[7$\times 10^{-7}$] T. 
The contribution to the external quadruopolar field from the charge-induced effect derived in this work thus becomes equal to the intrinsic contribution at an electron concentration of $\sim$10$^{7}$ per 1{mm} radius sphere, or $\sim$10$^{10}$ cm$^{-3}$. This concentration is 
likely achievable with field effect doping (although this might not be convenient in the spherical geometry).
For the larger sphere used in the measurements of Ref.~\onlinecite{Astrov:2013ua} the predicted value from the intrinsic mechanism is correspondingly smaller because of the $\frac{1}{V^{1/3}}$ dependence, and is indeed consistent with the measured value of $\vec{B}_{exp}=$\unit[1$\times 10^{-9}$]{T} \cite{Astrov:2013ua}.
The contribution from the charge effect becomes comparable at even smaller charging levels, and a superposition of the two contributions might be responsible for the complicated measured radial dependence.

Finally we point out that there is an additional contribution to the magnetic field around a sphere of 
antiferromagnetic material, resulting from truncation of the discrete spin arrangement of an antiferromagnet \cite{Andreev:1996vi}. These ``Andreev fields" are in principle distinguishable from the intrinsic and charge-induced magnetoelectric fields as they are genuine surface fields which decay exponentially with the distance from the surface. 
In addition, their form is sensitive to the details of the surface termination and is not required to be quadrupolar. Recently it was argued that the Andreev field is fundamentally different in a magnetoelectric material than in a non-magnetoelectric antiferromagnet\cite{Belashchenko:2010ky}.

\subsection{Conclusion}
In summary, we have derived the static electric and magnetic fields induced by a free charge in a diagonal magnetoelectric. We found that the electric fields are analogous to those generated by a charge in a simple dielectric, with a renormalization of the dielectric permittivity by the magnetoelectric response. A charge generates both monopolar and quadrupolar magnetic fields inside the material, stemming from the isotropic or tracefree parts of the ME tensor respectively. Consistent with Maxwell's equations, however, the monopolar magnetic contribution does not propagate outside of a finite sample, and so magnetoelectrics with the same tracefree part of the $\alpha$ tensor are indistinguishable by magnetic field measurements. Interestingly, even a sphere of purely monopolar isotropic magnetoelectric material will have an external quadrupolar magnetic field because of the interactions between dielectric and magnetic permeabilities and the magnetoelectric tensor. The magnitude of the external magnetic field induced by charges can be comparable with the intrinsic quadrupolar field as well as that from the surface magnetization for realistic geometries, and all components should be considered in interpreting measurements of fields around magnetoelectrics. 
\bibliography{biblio}{}

\newpage
\newpage

\appendix
\appendix
\section{Detailed derivation for a point charge in an isotropic magnetoelectric}\label{app:iso_inf}

Here we present the details of the derivation for the case of a point charge in an isotropic 
magnetoelectric material. We follow Ref.~\onlinecite{Jackson:2007ub} to calculate the scalar electric and magnetic potentials, $\phi_e(r)$ and $\phi_m(r)$, using the materials equations
\begin{eqnarray*}
\vec{B} & = & \mat{\mu}\vec{H}+\mat{\alpha} \vec{E}\\
\vec{D} & = & \mat{\epsilon}\vec{E}+\mat{\alpha} \vec{H} \quad,
\end{eqnarray*}
where the last terms in each expression result from the magnetoelectric effect, and the permittivity and permeability tensors $\mat{\mu}$ and $\mat{\epsilon}$ have the same symmetry as $\mat{\alpha}$. Using $\nabla \times \vec{H}=0$ and thus $\vec{H}=-\nabla\phi_m(\vec{r})$ (as there is no free charge current), $\nabla \cdot \vec{B}=0$ and $\nabla\cdot\vec{D}=\rho$, we obtain 
\begin{eqnarray}\label{eqn:def_phi_em}
\mat{\mu} \nabla^2 \phi_m(\vec{r})      & + & \mat{\alpha} \nabla^2 \phi_e(\vec{r})=0 \nonumber \\
\mat{\epsilon} \nabla^2 \phi_e(\vec{r}) & + & \mat{\alpha} \nabla^2 \phi_m(\vec{r})=q\delta(\vec{r}) \quad .
\end{eqnarray}

Solving Eqs.~\ref{eqn:def_phi_em} simultaneously we obtain trivially:
\begin{eqnarray*}
\phi_e(\vec{r})&=&\frac{q \mu}{4\pi(\epsilon\mu-\alpha^2)} \frac{1}{\vec{r}}\\
\phi_m(\vec{r})&=&-\frac{\alpha}{\mu}\phi_e(\vec{r}) =-\frac{q \alpha}{4\pi(\epsilon\mu-\alpha^2)} \frac{1}{\vec{r}} \quad .\\
\end{eqnarray*}
The fields are then obtained straightforwardly from the gradients of the potentials:
\begin{eqnarray*}
\vec{E}(\vec{r})&=&\frac{ \mu}{4\pi(\epsilon\mu-\alpha^2)} \frac{q }{\vec{r^2}}\vec{e_r}\\
 \vec{H}(\vec{r})&=&-\frac{ \alpha}{4\pi(\epsilon\mu-\alpha^2)} \frac{q}{\vec{r^2}}\vec{e_r} \quad,
\end{eqnarray*}
with $\vec{e_r}$ the radial unit vector.

\section{Detailed derivation for a point charge in a sphere of isotropic magnetoelectric in vacuum}\label{app:iso_boundary}
For the boundary problem we proceed by expanding the potentials $\phi_m(\vec{r})$ and $\phi_e(\vec{r})$ in terms of Legendre Polynomials, $P_l(cos(\theta))$:
\begin{equation}
\phi_i(\vec{r},\theta)=\sum_l (2l+1) \left[ A_l \vec{r}^l + B_l \vec{r}^{-(l+1)} \right]P_l(cos(\theta))
\end{equation}
with expansion coefficients $A_l$ and $B_l$. In the limit $\vec{r} \rightarrow \infty$ both potentials tend to
zero, and so $A_l=0$ for the outside potentials. For  $\vec{r} \rightarrow 0$ the potentials should not diverge thus $B_l$ is zero for the inside potentials. Therefore, using our results for the point charge in an infinite medium, we write the Ansatz: 
\begin{eqnarray}
\phi^i_e(\vec{r})&=&\frac{q \mu}{4\pi(\epsilon\mu-\alpha^2)} \frac{1}{\vec{r}}+\sum_l (2l+1)  A_l \vec{r}^l  P_l(cos(\theta))\nonumber \\
\phi^o_e(\vec{r})&=&\sum_l (2l+1)  B_l \vec{r}^{-(l+1)}  P_l(cos(\theta))\nonumber \\
\phi^i_m(\vec{r})&=&-\frac{q \alpha}{4\pi(\epsilon\mu-\alpha^2)} \frac{1}{\vec{r}}+\sum_l (2l+1)  C_l \vec{r}^l  P_l(cos(\theta))\nonumber \\
\phi^o_m(\vec{r})&=&\sum_l (2l+1)  D_l \vec{r}^{-(l+1)}  P_l(cos(\theta)) \quad,
\end{eqnarray}
where the $i$ and $o$ superscripts indicate the solutions for inside and outside of the sphere. (Since the Legendre expansion solves only the homogenous equation, the source terms must be added to account for the charge in the center.) The coefficients are found from the boundary conditions for the corresponding fields:
\begin{eqnarray}
\phi_{e,m}^i(\vec{R})&=&\phi_{e,m}^o(\vec{R})\nonumber \\
\vec{D}^i(\vec{R})\cdot \vec{n}&=&\vec{D}^o(\vec{R})\cdot \vec{n}\nonumber \\
\vec{B}^i(\vec{R})\cdot \vec{n}&=&\vec{B}^o(\vec{R})\cdot \vec{n} \quad,
\end{eqnarray}
where $\vec{n}$ is a unit vector in the radial direction.
Inserting $\vec{D}$ and $\vec{B}$ explicitly in the last two equations and changing to potentials rather then fields yields
\begin{eqnarray}
(\epsilon \vec{E}^i+\alpha \vec{H}^i)\cdot \vec{n}&=&(\epsilon_o \vec{E}^o) \cdot \vec{n}\nonumber \\
\epsilon \nabla_{\vec{r}} \phi^e_i(\vec{r})+\alpha \nabla_{\vec{r}} \phi^i_m(\vec{r})&=&\epsilon_o \nabla_{\vec{r}} \phi_e^o(\vec{r})\nonumber \\
(\mu \vec{H}^i+\alpha \vec{E}^i)\cdot \vec{n}&=&(\mu_o \vec{H}^o) \cdot \vec{n}\nonumber \\
\mu \nabla_{\vec{r}} \phi^m_i(\vec{r})+\alpha \nabla_{\vec{r}} \phi^i_e(\vec{r})&=&\mu_o \nabla_{\vec{r}} \phi_m^o(\vec{r}) \quad,
\end{eqnarray}
leading to the solution
\begin{eqnarray}
A_0&=&\dfrac{(\epsilon\mu)-\mu\epsilon_0}{4\pi (\epsilon\mu-\alpha^2) R}\nonumber \\
B_0&=&\dfrac{q}{4\pi \epsilon_0}\nonumber \\
C_0&=&\dfrac{\alpha}{4\pi(\epsilon\mu-\alpha^2)R}\nonumber \\
D_0&=&0 \quad ,
\end{eqnarray}
where $R$ is the sphere radius and all other coefficients are zero. 

\section{Detailed derivation for a point charge in a diagonal uniaxial magnetoelectric}\label{app:uniaxial_inf}
Now we consider the case of a uniaxial material with magnetoelectric tensor of the form:
\begin{eqnarray*}
\mat{\alpha}= \begin{bmatrix} \alpha_\perp&0&0\\ 0&\alpha_\perp&0 \\ 0&0&\alpha_\| \end{bmatrix} 
\end{eqnarray*}
Then instead of equations \ref{eqn:def_phi_em} we have:
\begin{widetext}
\begin{eqnarray}
\mu_\perp \nabla_\perp^2 \phi_m(\vec{r})+\mu_\| \nabla_\|^2 \phi_m(\vec{r})+\alpha_\perp \nabla_\perp^2 \phi_e(\vec{r})+\alpha_\| \nabla_\|^2 \phi_e(\vec{r})&=&0\nonumber \\
\epsilon_\perp \nabla_\perp^2 \phi_e(\vec{r})+\epsilon_\| \nabla_\|^2 \phi_e(\vec{r})+\alpha_\perp \nabla_\perp^2 \phi_m(\vec{r})+\alpha_\| \nabla_\|^2 \phi_m(\vec{r})&=&q\delta(\vec{r}) \quad,
\end{eqnarray}
Fourier transformation of the above expression yields
\begin{eqnarray}
\mu_\perp \vec{k}_\perp^2 \phi_m(\vec{k})+\mu_\| \vec{k}_\|^2 \phi_m(\vec{k})+\alpha_\perp \vec{k}_\perp^2 \phi_e(\vec{k})+\alpha_\| \vec{k}_\|^2 \phi_e(\vec{k})&=&0 \nonumber \\
\epsilon_\perp \vec{k}_\perp^2 \phi_e(\vec{k})+\epsilon_\| \vec{k}_\|^2 \phi_e(\vec{k})+\alpha_\perp \vec{k}_\perp^2 \phi_m(\vec{k})+\alpha_\| \vec{k}_\|^2 \phi_m(\vec{k})&=&q \quad .\label{eqn:phi_uniaxial1}
\end{eqnarray}
Eqns.~\ref{eqn:phi_uniaxial1} can then be solved to yield the following expressions for the scalar potentials in the uniaxial case:
\begin{eqnarray}
\phi_e(\vec{k})& = & \frac{\mu_\perp \vec{k}_\perp^2+\mu_\| \vec{k}_\|^2}{(\epsilon_\perp \vec{k}_\perp^2+\epsilon_\| \vec{k}_\|^2)(\mu_\perp \vec{k}_\perp^2+\mu_\| \vec{k}_\|^2)-(\alpha_\perp \vec{k}_\perp^2 +\alpha_\| \vec{k}_\|^2)^2}q \nonumber \\
\phi_m(k) & = & -1\frac{\alpha_\perp k_\perp^2 +\alpha_\| k_\|^2 }{\mu_\perp k_\perp^2+\mu_\| k_\|^2}\phi_e(k) \quad .
\end{eqnarray}
Transforming to spherical coordinates $k_\perp=k\,sin(\theta)$ and $k_\|=k\,cos(\theta)$ yields 
\begin{equation}
\phi_e(k)=\frac{\mu_\perp sin(\theta)^2+\mu_\| cos(\theta)^2}{(\epsilon_\perp sin(\theta)^2+\epsilon_\| cos(\theta)^2)(\mu_\perp sin(\theta)^2+\mu_\| cos(\theta)^2)-(\alpha_\perp sin(\theta)^2 +\alpha_\| cos(\theta)^2)^2}\frac{q}{k^2} =
f_e({\theta})\frac{q}{k^2} \quad .
\end{equation}
\end{widetext}
The real space solution is then obtained from the inverse transformation
\begin{equation*}
\phi_e(\vec{r})=\int_0^{2\pi} \int_0^\pi \int_0^\infty f_e({\theta})\frac{q}{k^2} e^{i \vec{k}\cdot\vec{r}} sin(\theta) k^2 d\phi\, d\theta\,  dk,
\end{equation*}
where by symmetry 
\begin{equation*}
\vec{k}\cdot\vec{r}=cos(\theta)cos(\theta')+sin(\theta)sin(\theta')cos(\phi-\phi')  \quad .
\end{equation*}
Setting $\phi-\phi'=\phi$, the integration over $\phi$ gives
\begin{equation*}
\frac{1}{r}\int_0^{2\pi}e^{ik~sin(\theta)sin(\theta')cos(\phi)}d\phi=2\pi J_0(ksin(\theta)sin(\theta')),
\end{equation*}
with $J_0(x)$ the Bessel function, and that over $k$ is the standard tabulated integral
\begin{widetext}
\begin{equation*}
\frac{1}{\vec{r}}\int_0^{\infty} J_0(ksin(\theta)sin(\theta'))e^{ik~cos(\theta)cos(\theta')}dk
 = \xi(\theta,\theta')= \begin{cases} 0 &\textnormal{for}~~~ cos(\theta)^2cos(\theta')^2>sin(\theta)^2sin(\theta')^2\\
(\sqrt{cos(\theta)^2cos(\theta')^2-sin(\theta)^2sin(\theta')})^{-1} &\textnormal{else}
\end{cases}
\end{equation*}
\end{widetext}
so we arrive at the final expression for $\phi_e$:
\begin{equation}
\phi_e(r,\theta')=\frac{2\pi q}{r}\int_0^\pi f_e(\theta)\xi(\theta,\theta')sin(\theta)d\theta \quad .
\end{equation}
Using the same approach we obtain 
\begin{equation}
\phi_m(r,\theta')=\frac{2\pi q}{r}\int_0^\pi f_m(\theta)\xi(\theta,\theta')sin(\theta)d\theta
\end{equation}
with
\begin{widetext}
\begin{equation}
f_m(\theta)=\frac{\alpha_\perp sin(\theta)^2+\alpha_\| cos(\theta)^2}{(\epsilon_\perp sin(\theta)^2+\epsilon_\| cos(\theta)^2)(\mu_\perp sin(\theta)^2+\mu_\| cos(\theta)^2)-(\alpha_\perp sin(\theta)^2 +\alpha_\| cos(\theta)^2)^2} \quad .
\end{equation}
\end{widetext}
We can rewrite the integral equations in terms of the function $\xi(\theta,\theta')$  as
\begin{equation}
\phi_m(r,\theta')=\frac{2\pi q}{r}\int_{|\pi/2-\theta'|}^{\pi-|\pi/2-\theta'|} f_m(\theta)\xi_2(\theta,\theta')sin(\theta)d\theta\;,
\end{equation}
with 
\begin{equation}
\xi_2(\theta,\theta')=\left(\sqrt{cos(\theta)^2cos(\theta')^2-sin(\theta)^2sin(\theta')}\right)^{-1}
\end{equation}

\section{Detailed derivation for a point charge in a diagonal uniaxial magnetoelectric sphere}\label{app:uniaxial_boundary}

For the uniaxial sphere we start from the previous considerations and expand the two potentials again in Legendre Polynoms:
\begin{eqnarray}
\phi_e(r,\theta)& = &\frac{1}{r}\sum_{n=1}^\infty C^e_{2n} P_{2n}(cos(\theta))\nonumber \\
\phi_m(r,\theta)& = &\frac{1}{r}\sum_{n=1}^\infty C^m_{2n} P_{2n}(cos(\theta)),
\end{eqnarray}
where the coefficients $C^e$ and $C^m$ arise from the solutions of the bulk uniaxial case. Due to symmetry only even terms appear in this expansion. Furthermore for physically reasonable parameters the coefficients converge rapidly and terms larger than $2n=8$ are practically zero so the series can be truncated in this order. 

Next we consider the boundary problem where we solve again the following system of equations:
\begin{eqnarray}
\phi_{e,m}^i(\vec{R})&=&\phi_{e,m}^o(\vec{R})\nonumber \\
\vec{D}^i(\vec{R})\cdot \vec{n}&=&\vec{D}^o(\vec{R})\cdot \vec{n}\nonumber \\
\vec{B}^i(\vec{R})\cdot \vec{n}&=&\vec{B}^o(\vec{R})\cdot \vec{n}\nonumber \\
\vec{E}^i(\vec{R})\cdot \vec{t}&=&\vec{E}^o(\vec{R})\cdot \vec{t}\nonumber \\
\vec{H}^i(\vec{R})\cdot \vec{t}&=&\vec{H}^o(\vec{R})\cdot \vec{t}\; \quad ,
\end{eqnarray}
where $\vec{t}$ is a unit vector in the tangential direction.
Here $i$ and $o$ superscript denote the solutions inside and outside of the sphere. The potentials are given by the following equations:
\begin{eqnarray}
\phi^i_e(\vec{r})&=&\sum_l \left( \frac{C^e_{l}}{r} + (2l+1)\;A_l \vec{r}^l \right) P_l(cos(\theta))\nonumber \\
\phi^o_e(\vec{r})&=&\sum_l (2l+1)  \frac{B_l}{\vec{r}^{-(l+1)}}  P_l(cos(\theta))\nonumber \\
\phi^i_m(\vec{r})&=&\sum_l \left( \frac{C^m_{l}}{r} + (2l+1)\,C_l \vec{r}^l \right) P_l(cos(\theta))\nonumber \\
\phi^o_m(\vec{r})&=&\sum_l (2l+1)  \frac{D_l}{\vec{r}^{-(l+1)}}  P_l(cos(\theta)).
\end{eqnarray}
Again we superimposed here our solutions with those of the homogenous Laplace equations, taking into account the limits $r\rightarrow0$ and $r\rightarrow\infty$, for inside and outside of the sphere, respectively. The four sets of coefficients, $A_l, B_l, C_l$ and $D_l$, are nonzero only for $l=2n$. As before, the displacement field $\vec{D}^i$ and magnetic induction $\vec{B}^i$ inside the sphere are given by:
\begin{eqnarray*}
\vec{D}^i &=& \epsilon \vec{E}+\alpha \vec{H} \\
\vec{B}^i &=& \mu \vec{H} + \alpha \vec{E}
\end{eqnarray*}
and outside the sphere by the usual expressions for the vacuum
\begin{eqnarray*}
\vec{D}^o &=& \vec{E}\\
\vec{B}^o &=& \vec{H} .
\end{eqnarray*}
Since we are working in spherical coordinates, we have to transform our tensors to the spherical coordinate system. Thus $\alpha$ in cartesaian:
\begin{equation}
\mat{\alpha}_{cart}= \begin{bmatrix} \alpha_\perp&0&0\\ 0&\alpha_\perp&0 \\ 0&0&\alpha_\| \end{bmatrix}
\end{equation}
becomes in spherical coordinates
\begin{equation}
\mat{\alpha}_{sph}= \begin{bmatrix} \alpha_{11}&\alpha_{12}&0\\ \alpha_{21}&\alpha_{22}&0 \\ 0&0&\alpha_{33} \end{bmatrix},
\end{equation}
with the four coefficients given by
\begin{eqnarray}
\alpha_{11}&=&\frac{1}{2}\left[(\alpha_\perp +\alpha_\|)+(\alpha_\|-\alpha_\perp)cos(2\theta)  \right]\nonumber \\
\alpha_{12}&=&\alpha_{21}=\left[(\alpha_\perp -\alpha_\|)\right]sin(\theta)cos(\theta)\nonumber \\
\alpha_{22}&=&\frac{1}{2}\left[(\alpha_\perp +\alpha_\|)+((\alpha_\perp -\alpha_\|)cos(2\theta)  \right]\nonumber \\
\alpha_{33}&=&\alpha_\perp\;.
\end{eqnarray}
For the radial contributions at the sphere boundary only $\alpha_{11}$ and $\alpha_{12}$ enter and the normal component of $\vec{D}^i$ becomes: 
\begin{equation}
\vec{D}^i\cdot \vec{n}=\epsilon_{11}\frac{\partial \phi^i_e}{\partial r} + \epsilon_{12} \frac{1}{r}\frac{\partial \phi^i_e}{\partial \theta}+\alpha_{11}\frac{\partial \phi^i_m}{\partial r} + \alpha_{12} \frac{1}{r}\frac{\partial \phi^i_m}{\partial \theta}
\end{equation}
and $\vec{D}^o$:
\begin{equation}
\vec{D}^o\cdot \vec{n}=\frac{\partial \phi^o_e}{\partial r}.
\end{equation}
For the tangential component of the electric field we have the following equations
\begin{equation}
\vec{E}^i\cdot \vec{t}=\frac{1}{r}\frac{\partial \phi^i_e}{\partial \theta}
\end{equation}
and 
\begin{equation}
\vec{E}^o\cdot \vec{t}=\frac{1}{r}\frac{\partial \phi^o_e}{\partial \theta}  \quad .
\end{equation}
Finally, to obtain the system of equations for the coefficients we multiply by $P_n(cos(\theta))$ (for normalization of the Legendre Polynomials= and integrate over $\frac{2n+1}{n}\int_0^\pi d\theta$ We obtain that all odd coefficients are zero, and that the coefficients $B_l$ and $D_l$ decay as expected with increasing $l$. 
\end{document}